\documentclass[conference]{IEEEtran}
\IEEEoverridecommandlockouts
\usepackage{amsmath,amssymb,amsfonts}
\usepackage{algorithmic}
\usepackage{graphicx}
\usepackage{textcomp}
\usepackage{fancyhdr}
\usepackage{xcolor}
\usepackage{lipsum}
\usepackage{enumitem} 
\usepackage{algorithm}
\usepackage{algorithmic}
\usepackage[sorting = none, backend = bibtex, style=numeric-comp]{biblatex}
\def\BibTeX{{\rm B\kern-.05em{\sc i\kern-.025em b}\kern-.08em
    T\kern-.1667em\lower.7ex\hbox{E}\kern-.125emX}}

\begin{document}




    

\clearpage
\thispagestyle{empty}

\clearpage
\onecolumn
\thispagestyle{empty}
\begin{center}
\vspace*{\fill}
\large
© 2026 IEEE. Personal use of this material is permitted. Permission from IEEE must be obtained for all other uses, in any current or future media, including reprinting/republishing this material for advertising or promotional purposes, creating new collective works, for resale or redistribution to servers or lists, or reuse of any copyrighted component of this work in other works.
\vspace*{\fill}
\end{center}
\newpage
\twocolumn
\setcounter{page}{1}

\title{A Hierarchical Framework for Graph Structure Learning in Histopathology Image Classification}

\author{
    \IEEEauthorblockN{Sudipta Paul}
    \IEEEauthorblockA{\textit{Department of Electrical Engineering} \\
    \textit{Rensselaer Polytechnic Institute}\\
    Troy, New York, USA \\
    pauls5@rpi.edu}
    \and
    \IEEEauthorblockN{Amanda W. Lund}
    \IEEEauthorblockA{\textit{Department of Pathology} \\
    \textit{NYU Grossman School of Medicine}\\
    New York, NY, USA \\
    amanda.lund@nyulangone.org}
    \and
    \IEEEauthorblockN{B\"{u}lent Yener}
    \IEEEauthorblockA{\textit{Department of Computer Science} \\
    \textit{Rensselaer Polytechnic Institute}\\
    Troy, New York, USA \\
    yener@cs.rpi.edu}
}

\maketitle

\begin{abstract}
The spatial organization of cells and tissues provides important diagnostic cues in histopathology images. Although graph-based approaches can model these relationships, many rely on fixed or heuristic graph structures that may not accurately represent tissue connectivity. In this work, we propose $G_2^*$-Net, an optimized two-level graph learning framework for classifying large-scale histopathology images, such as whole-slide images (WSIs) or large regions of interest (ROIs). Here, $G_2$ denotes the two-level hierarchical graph representation, and the superscript $*$ indicates the optimized image-level graph structure learned from the proposed framework. The method first divides each WSI or large ROI into image patches, constructs cell-level graphs within each patch to capture local tissue architecture, and then represents each patch as a node in a learnable image-level graph. $G_2^*$-Net formulates image-level graph structure learning as a second-order bilevel optimization problem, separating graph connectivity learning from classifier optimization while coupling them through validation-driven feedback. To make this formulation computationally practical, we adopt a DARTS-inspired one-step unrolled approximation for efficient hypergradient estimation. Experimental validation on three distinct histopathology datasets demonstrates the effectiveness of our proposed method.
\end{abstract}

\section{Introduction}

Deep learning has transformed histopathology image analysis by enabling data-driven extraction of complex visual patterns from tissue images. Convolutional neural networks (CNNs), in particular, have achieved strong performance compared to conventional machine learning approaches~\cite{cnn1_review, extend, china}. In CNN-based pipelines, whole-slide images (WSIs) or large regions of interest (ROIs) are divided into smaller patches, and patch-level predictions are aggregated into an image-level decision. Although this strategy makes high-resolution images computationally manageable, independent patches may miss long-range tissue relationships and include irrelevant regions. To capture broader tissue organization, graph convolutional network (GCN)-based methods using cell graph representations have been increasingly explored~\cite{cg,cgcnet,hatnet,c2pgcn}. These methods model spatial and topological relationships among cells, but many rely on fixed graphs with predefined connections that may introduce inductive biases and inadequately represent context-dependent tissue interactions.

To reduce reliance on heuristic connectivity rules, adaptive graph structure learning methods have gained increasing attention, as they update graph connectivity during training rather than relying on fixed edges. Existing approaches dynamically refine adjacency matrices through learnable transformation functions~\cite{adnan,liu}, jointly learn graph construction using CNN-based filters~\cite{cnngraphlearn1,cnngraphlearn2}, or select representative patches as slide-level graph nodes~\cite{kim,patch2}. However, most methods jointly optimize graph structure and node representations within a single-level objective, providing no explicit validation-driven mechanism for learning task-specific connectivity.

To alleviate this issue, we introduce $G_2^*$-Net, an optimized two-level graph-based framework for histopathology image classification. Here, $G_2$ denotes a two-level graph hierarchy composed of local patch-level graphs and a global image-level graph, while the superscript $*$ indicates the optimized image-level structure learned by the framework. $G_2^*$-Net partitions WSIs or large ROIs into overlapping patches, constructs cell graphs to capture local tissue architecture, and represents each patch descriptor as a node in a learnable image-level graph. In the current framework, the local patch-level representation is kept fixed so that the optimization focuses on image-level graph structure while maintaining computational tractability.

Our method formulates image-level graph structure learning as a second-order bilevel optimization problem. This separates graph connectivity learning from classifier parameter optimization while coupling them through a validation-driven bilevel objective. The second-order bilevel formulation, however, is computationally prohibitive because it requires repeatedly solving the lower-level classifier problem and computing Hessian-related second-order derivatives. We therefore adopt a DARTS-inspired one-step unrolled approximation~\cite{darts}, in which a virtual classifier update is performed using the training objective, followed by a graph-structure update using the validation objective. This enables efficient estimation of second-order hypergradients without solving the lower-level problem exactly. The key contributions of our work are:

\begin{itemize}[leftmargin=*]

\item We introduce a second-order bilevel optimization framework for image-level graph structure learning in histopathology image classification. The formulation separates graph connectivity learning from classifier optimization while explicitly incorporating second-order hypergradient information.

\item We design a two-level hierarchical graph representation that connects local patch-level cellular representations with global image-level graph optimization. Within this hierarchy, we use a DARTS-style one-step unrolled update \cite{darts} to efficiently approximate second-order bilevel optimization without solving the lower-level problem to convergence.

\item We conduct comprehensive experiments on three histopathology datasets: the Extended CRC dataset \cite{extend}, the Colon cancer dataset~\cite{china}, and the Melanoma dataset~\cite{melanoma}—demonstrating that the proposed method achieves performance on par with or exceeding that of CNN- and GCN-based baseline methods.
\end{itemize}

\section{Related Work}

\subsection{Cell Graph Representation for Histopathology Analysis}

Cell graphs model tissue organization by representing cells as nodes and their spatial relationships as edges, commonly defined using neighborhood or probabilistic criteria~\cite{cg,hatnet,cgcnet,hact}. Their ability to preserve cellular morphology and tissue architecture has motivated their increasing use in digital pathology.

Several methods combine cell graphs with graph neural networks. CGC-Net~\cite{cgcnet} uses multi-scale graph aggregation for colorectal cancer grading, while HAT-Net~\cite{hatnet} integrates graph learning, MinCut pooling, and Transformer modules to capture hierarchical tissue relationships. HACT-Net~\cite{hact} models cellular and tissue structures through a hierarchical cell-to-tissue graph, and C2P-GCN~\cite{c2pgcn} constructs local patch-level and global image-level graphs for colorectal cancer grading. These studies demonstrate the effectiveness of hierarchical graph representations for modeling tissue organization.

\subsection{Adaptive Graph Learning in Histology Images}

To address the limitations of heuristic graphs in histopathology analysis, recent studies have explored adaptive graph structure learning. For example, authors in~\cite{adnan} modeled WSIs as fully connected graphs of representative patches and learned graph connectivity dynamically for lung cancer subtype classification. 
The approach in~\cite{liu} updates sparse cosine-similarity-based adjacency matrices for survival prediction, while~\cite{li} generates edges from learned head--tail patch embeddings using knowledge-aware attention. Another approach selects representative
microscopy images as graph nodes through clustering and
Gumbel-Softmax sampling~\cite{kim}.

Despite their effectiveness, most existing methods jointly optimize graph construction and prediction within a single-level objective, limiting the explicit separation of structure and representation learning. A first-order approximate bilevel method was recently introduced for histopathology graph learning~\cite{FOA}; however, it neglects the dependence of the optimized classifier parameters on the graph-structure parameters. We instead formulate image-level graph learning as a second-order
bilevel problem, optimizing graph connectivity at the upper level
and classifier parameters at the lower level. The resulting
hypergradient captures the classifier's response to connectivity changes.

\section{Methodology}


\subsection{Cell Identification}

After each WSI or ROI is divided into patches, nuclei are detected within each patch to obtain cell-level spatial locations. This step provides the cell centroids required for constructing patch-level graphs. We use a pretrained StarDist2D model \cite{stardist} for nuclei detection in this work. The detected nuclear centroids are then used as graph nodes, enabling spatial graph construction at the patch level.

\begin{figure}[h]
\centering
\includegraphics [width=8cm, height=2cm]{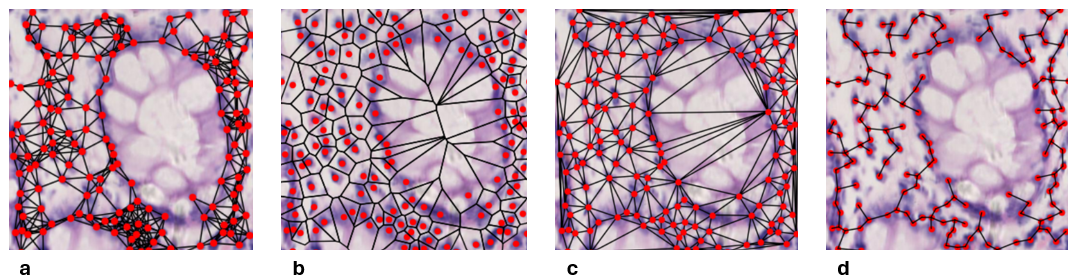}
\caption{Visualization of (a) cell graph constructed using Euclidean distance, $r_p = 64$, (b) Voronoi diagram, (c) Delaunay triangulation, and (d) Minimum spanning tree on a small portion of a patch.}
\label{fig:figure_2}
\end{figure}

\subsection{Patch-level Graph}

Within each image patch $p$, the nuclei form a cell graph ${G}_p=({V}_p,{E}_p)$ representing local cellular organization, where ${V}_p$ denotes the identified cells and ${E}_p$ their spatial connections. Since cell-to-cell interactions cannot be reliably determined from the image alone, graph connectivity is defined with a proximity rule. Specifically, two cells $i$ and $j$ are connected if their Euclidean distance $\delta_p(i,j)$ is smaller than a predefined radius $r_p$. The adjacency matrix ${A}_p$ is therefore defined as
\begin{equation}
{A}_p(i,j)=
\begin{cases}
1, & \text{if } \delta_p(i,j) < r_p,\\
0, & \text{otherwise}.
\end{cases}
\end{equation}

The resulting graph summarizes local tissue organization, including cell clustering, crowding, and isolation. An example of the patch-level cell graph is shown in Fig.~\ref{fig:figure_2}(a). We then extract $18$ graph-based descriptors from each patch to quantify these local spatial patterns, as summarized in Table~\ref{tab:patch_graph_features}~\cite{c2pgcn}.

\begin{table}[!h]
\caption{Local neighborhood cell graph features}
\label{tab:patch_graph_features}
\centering
\begin{tabular}{@{}|p{2cm}|c|p{5cm}|@{}}
\hline
\textbf{Feature type} & \textbf{No.} & \textbf{Description} \\ \hline
Connectedness and cliquishness measures & 4 & Clustering coefficient, Average degree, Number of connected components, Giant connected component ratio \\ \hline
Distance-based measures & 8 & Number of vertices, Number of edges, Radius, Diameter, Average eccentricity, Number of central points, Percent of central points, Average path length \\ \hline
Spectral Measures & 6 & Trace of adjacency, Energy of adjacency, Trace of Laplacian, Lower slope, Upper slope, Largest eigenvalue adjacency \\ \hline
\end{tabular}
\end{table}

\begin{figure*}[t]
\centering
\includegraphics[width=13cm, height=5.2cm]{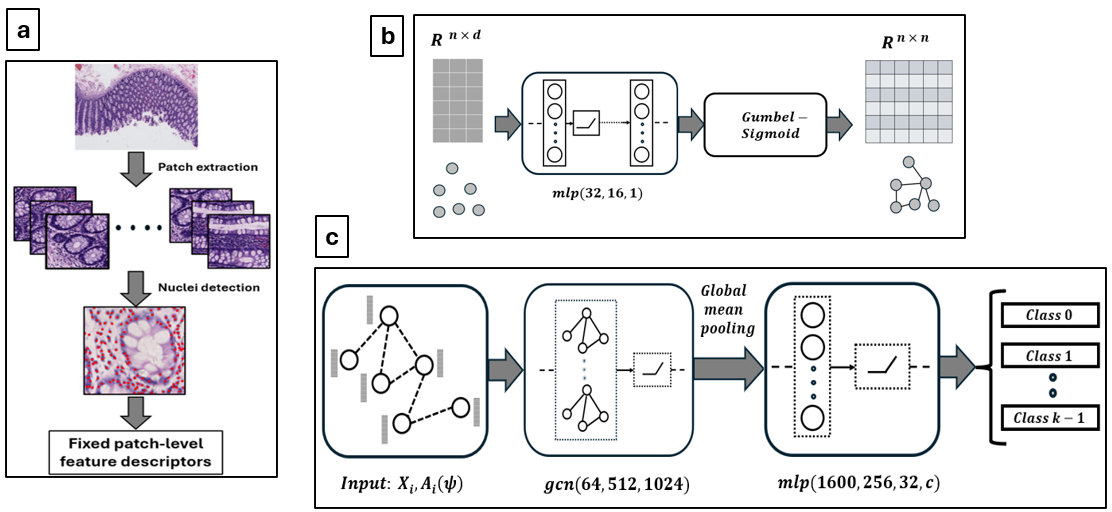}
\caption{Overview of $G_2^*$-Net. (a) Patch-level representation: WSIs or large ROIs are divided into overlapping patches, from which cell graphs, Voronoi diagrams, Delaunay triangulations, and minimum spanning trees provide fixed structural descriptors. (b) Adjacency generation: patch descriptors form graph nodes, and an MLP with
Gumbel-Sigmoid learns pairwise connections. (c) Image-level classification: the learned adjacency and node features are processed by GCN layers.
Multi-layer embeddings are globally mean-pooled, concatenated using a Jumping Knowledge-style readout, and classified by an MLP head.}
\label{fig:your_label}
\end{figure*}

In addition to the local neighborhood cell graph features, we organize the patch-level structural descriptors into several feature families that capture complementary aspects of tissue architecture. These include Voronoi Diagram (VD), Delaunay Triangulation (DT), Minimum Spanning Tree (MST), and nuclear Nearest-Neighbor (NN) feature families. Representative VD, DT, and MST examples are shown in Fig.~\ref{fig:figure_2}(b), Fig.~\ref{fig:figure_2}(c), and Fig.~\ref{fig:figure_2}(d), respectively. We compute $24$ structural descriptors, comprising $12$ VD, $8$ DT, and $4$ MST features, together with $27$ NN features characterizing cellular aggregation. These descriptors are summarized in Table~\ref{tab:table_example}~\cite{c2pgcn}, and the overall patch-level graph construction is illustrated in Fig.~\ref{fig:your_label}(a).

\begin{table}[!h]
\caption{Patch-level Global cell graph features}
\label{tab:table_example}
\centering
\begin{tabular}{@{}|p{1.5cm}|c|p{5.5cm}|@{}}
\hline
\textbf{Feature type} & \textbf{No.} & \textbf{Description} \\ \hline
Voronoi diagram & 12 & Mean, SD, Min/Max Ratio, and Disorder of Chord length, Polygon area, and Perimeter.  \\ \hline
Delaunay triangulation & 8 & Mean, SD, Min/Max Ratio, and Disorder of Triangle side length and Area.  \\ \hline
Minimum spanning tree & 4 & Mean, SD, Min./Max. Ratio and Disorder of Edge length \\ \hline
Nuclei nearest neighbor (NN) features & 27 & Number of nuclei, area of polygons, density of nuclei; Mean, SD, and Disorder of distance to $k$-NN, where $k=3,5, 7$; Mean, SD, and disorder of NN within an $n$ pixel radius, where, $n = 10,20,30,40,50$ \\ \hline
\end{tabular}
\end{table}

Combining all feature families yields a $69$-dimensional vector
for each patch. Let ${f}_q^{(i)}\in\mathbb{R}^{d}$ denote the
feature vector of the $q$-th patch in image $i$, where $d=69$.
For an image with $n_i$ patches, the patch-level feature matrix
is defined as
\begin{equation}
{X}_i =
\left[
{f}_1^{(i)}, {f}_2^{(i)}, \ldots, {f}_{n_i}^{(i)}
\right]^{\top}
\in \mathbb{R}^{n_i \times d}.
\end{equation}

\subsection{Image-level Graph}

We construct an image-level graph to model patch interactions within each WSI or large ROI, treating each row of ${X}_i$ as a node. For image $i$, the image-level graph is denoted as $G_i^{\mathrm{img}}(\psi)=(V_i,E_i(\psi))$, where $V_i$ is the set of patch nodes, $E_i(\psi)$ denotes their learned connections, and $\psi$ denotes the trainable parameters. The graph connectivity is encoded by $A_i(\psi)\in \mathbb{R}^{n_i\times n_i}$, whose entries indicate pairwise connections.

Given $A_i(\psi)$, a GCN-based classifier $f_{\theta}$ produces the image-level prediction $\hat{y}_i=f_{\theta}(X_i,A_i(\psi))$, where $\theta$ denotes the classifier parameters. We formulate graph-structure and classifier learning as a bilevel problem, with $\psi$ and $\theta$ treated as the upper- and lower-level parameters, respectively. Let $\mathcal{L}_{\mathrm{train}}$ and $\mathcal{L}_{\mathrm{val}}$ denote the cross-entropy losses on the training and validation sets, respectively. The bilevel objective is written as
\begin{equation}
\begin{aligned}
    \underset{\psi}{\min}\; \mathcal{L}_{\text{val}}\left(\theta^*(\psi), \psi\right)
\end{aligned}
\end{equation}
\begin{equation}
\begin{aligned}
    \text{s.t.} \quad \theta^*(\psi) &= \underset{\theta}{\arg\min}\; \mathcal{L}_{\text{train}}(\theta, \psi).
\end{aligned}
\end{equation}

The exact bilevel optimization is computationally expensive because
each adjacency-generator update requires solving the lower-level
classifier problem to obtain $\theta^{*}(\psi)$. Moreover, the
upper-level update must account for the dependence of
$\theta^{*}(\psi)$ on $\psi$, introducing Hessian-related
second-order terms. For large GCN-based classifiers, repeatedly solving the inner problem and directly computing Hessian-related terms is impractical. We therefore adopt a DARTS-style one-step unrolled approximation~\cite{darts}. At iteration $t$, given the current parameters $\theta_{t-1}$ and $\psi_{t-1}$, we compute a temporary classifier update as
\begin{equation}
\tilde{\theta}_{t}
=
\theta_{t-1}
-
\xi
\nabla_{\theta}
\mathcal{L}_{\text{train}}(\theta_{t-1},\psi_{t-1}),
\end{equation}
where $\xi$ is the unrolled inner-step learning rate. The temporary parameter $\tilde{\theta}_{t}$ approximates how the classifier would change after one training step under the current learned graph structure.

The adjacency generator is then updated using the validation loss evaluated at $\tilde{\theta}_{t}$. We first define the direct validation gradient with respect to $\psi$ as
\begin{equation}
d_{\psi}^{(t)}
=
\nabla_{\psi}
\mathcal{L}_{\text{val}}(\tilde{\theta}_{t},\psi_{t-1}).
\end{equation}
This term captures the direct effect of the adjacency generator on the validation loss. To account for the indirect effect of $\psi$ through the one-step classifier
update, we also compute the validation gradient with respect to $\tilde{\theta}_t$:
\begin{equation}
v_t
=
\nabla_{\tilde{\theta}_{t}}
\mathcal{L}_{\text{val}}(\tilde{\theta}_{t},\psi_{t-1}).
\end{equation}
The second-order correction term is then written as the mixed Hessian-vector product
\begin{equation}
h_{\psi}^{(t)}
=
\nabla^2_{\psi,\theta}
\mathcal{L}_{\text{train}}(\theta_{t-1},\psi_{t-1})
v_t.
\end{equation}

Directly computing this product is expensive, so we approximate it using finite differences. Let $\epsilon_t=\frac{r}{\|v_t\|_2}$, where $r$ is a small radius. The classifier parameters are
then perturbed as
\begin{equation}
\theta_t^+
=
\theta_{t-1}
+
\epsilon_t v_t,
\qquad
\theta_t^-
=
\theta_{t-1}
-
\epsilon_t v_t.
\end{equation}
The Hessian-vector product is then approximated as \cite{darts}
\begin{equation}
\label{eq:mixed_hvp}
h_{\psi}^{(t)}
\approx
\frac{
\nabla_{\psi}
\mathcal{L}_{\text{train}}(\theta_t^+,\psi_{t-1})
-
\nabla_{\psi}
\mathcal{L}_{\text{train}}(\theta_t^-,\psi_{t-1})
}
{2\epsilon_t}.
\end{equation}

The finite-difference second-order correction may be unstable
and distort the direct validation-gradient update. We therefore introduce a reliability coefficient $\rho_t$, which adaptively controls the second-order correction. Defining the correction direction as $c_{\psi}^{(t)} = -\xi h_{\psi}^{(t)}$, we measure its agreement with the direct validation gradient using cosine similarity:
\begin{equation}
\gamma_t
=
\frac{
\left\langle d_{\psi}^{(t)}, c_{\psi}^{(t)} \right\rangle
}
{
\left\|d_{\psi}^{(t)}\right\|_2
\left\|c_{\psi}^{(t)}\right\|_2
+
\epsilon_0
},
\end{equation}
where $\epsilon_0$ is a small constant used for numerical stability. The
reliability coefficient is then defined as
\begin{equation}
\label{eq:rho_gamma}
\rho_t
=
\frac{1+\gamma_t}{2}.
\end{equation}
A larger $\rho_t$ assigns greater weight to an aligned correction, whereas a smaller value suppresses a poorly aligned correction. The resulting second-order hypergradient is
\begin{equation}
g_{\psi}^{(t)}
=
d_{\psi}^{(t)}
-
\rho_t\xi h_{\psi}^{(t)}.
\end{equation}
After that, the adjacency generator is updated
as
\begin{equation}
\psi_t
=
\psi_{t-1}
-
\eta_{\psi}
g_{\psi}^{(t)}.
\end{equation}
The classifier parameters are subsequently updated as
\begin{equation}
\theta_t
=
\theta_{t-1}
-
\eta_{\theta}
\nabla_{\theta}
\mathcal{L}_{\text{train}}(\theta_{t-1},\psi_t).
\end{equation}

The temporary parameters $\tilde{\theta}_{t}$ are discarded after the generator update. This procedure approximates the second-order bilevel update without fully solving the inner problem or explicitly forming the Hessian matrix.

In this study, the classifier $f_{\theta}$ is implemented as a GCN-based classification network with a Jumping Knowledge-style readout and an MLP head. Given $A_i(\psi)$, the GCN layers process patch-level node features to learn hierarchical representations. Embeddings from multiple layers are globally pooled and concatenated into an image-level representation, which is passed to the MLP head for classification. The detailed architecture of the network is illustrated in Fig.~\ref{fig:your_label}(c).

The image-level adjacency matrix $A_{i}(\psi)$ is generated by a trainable parametric function. Specifically, we define
\begin{equation}
q_{\psi}(X_i): \mathbb{R}^{n_i \times d} \rightarrow \mathbb{R}^{n_i \times n_i},
\end{equation}
where $q_{\psi}$ is implemented using an MLP as shown in Fig.~\ref{fig:your_label}(b) and produces an edge-score matrix $S_i \in \mathbb{R}^{n_i \times n_i}$.
For each patch pair $(m,n)$, their feature vectors are concatenated and passed through the MLP to produce a scalar connectivity logit:
\begin{equation}
S_{i}(m, n)
=
q_{\psi}(f_m^{(i)}, f_n^{(i)}).
\end{equation}

To obtain differentiable stochastic adjacency values, we use
Gumbel-Sigmoid reparameterization~\cite{gumbel}:
\begin{equation}
A_i(\psi)_{m,n}
=
\sigma
\left(
\frac{
S_{i}(m,n)+g_{m,n}
}{\tau}
\right),
\end{equation}
where $g_{m,n}\sim\mathrm{Gumbel}(0,1)$ is the sampled Gumbel noise, $\sigma(\cdot)$ denotes the sigmoid function and $\tau$ is the temperature. We anneal the temperature as
$\tau^{(t)}=\max\left(\tau_{\min},
\alpha_{\tau}\tau^{(t-1)}\right)$, where $\tau_{\min}$ is its
lower bound and $\alpha_{\tau}\in(0,1)$ is the decay factor. Since patch interactions are undirected, we symmetrize the
adjacency matrix as
$A_i(\psi)=\left(A_i(\psi)+A_i(\psi)^T\right)/2$
and add self-connections. The resulting matrix is then used by the GCN-based classifier to perform image-level classification. The overall learning process is summarized in Algorithm 1.

\begin{algorithm}[h]
\caption{Image-level Training Procedure}
\label{alg:$G_2^*$-Net}
\begin{algorithmic}[1]
\REQUIRE $\mathcal{D}_{\mathrm{train}}$, $\mathcal{D}_{\mathrm{val}}$, learning rates $\eta_{\theta}, \eta_{\psi}$, unrolled step size $\xi$, finite-difference radius $r$, temperature $\tau_{\mathrm{init}}, \tau_{\mathrm{min}}, \alpha_\tau$
\ENSURE Classifier parameters $\theta^{*}$, adjacency parameters $\psi^{*}$

\STATE Initialize parameters $\theta_{0}$, $\psi_{0}$, and $\tau \leftarrow \tau_{\mathrm{init}}$
\STATE $t \leftarrow 1$

\WHILE{not converged}

    \STATE \textbf{One-step unrolled classifier update:}
    \STATE $\tilde{\theta}_{t} \leftarrow \theta_{t-1} - \xi \nabla_{\theta}\mathcal{L}_{\mathrm{train}}(\theta_{t-1},\psi_{t-1})$

    \STATE \textbf{Second-order adjacency generator update:}
    \STATE $d_{\psi}^{(t)} \leftarrow \nabla_{\psi}\mathcal{L}_{\mathrm{val}}(\tilde{\theta}_{t},\psi_{t-1})$
    \STATE $v_t \leftarrow \nabla_{\tilde{\theta}_{t}}\mathcal{L}_{\mathrm{val}}(\tilde{\theta}_{t},\psi_{t-1})$
    \STATE Approximate $h_{\psi}^{(t)}$ using \eqref{eq:mixed_hvp}
    \STATE Compute reliability coefficient $\rho_t$ as~\eqref{eq:rho_gamma}.
    \STATE $g_{\psi}^{(t)} \leftarrow d_{\psi}^{(t)} - \rho_t \xi h_{\psi}^{(t)}$
    \STATE $\psi_t \leftarrow \psi_{t-1} - \eta_{\psi} g_{\psi}^{(t)}$

    \STATE \textbf{Actual classifier update:}
    \STATE $\theta_t \leftarrow \theta_{t-1} - \eta_{\theta}\nabla_{\theta}\mathcal{L}_{\mathrm{train}}(\theta_{t-1},\psi_t)$

    \STATE \textbf{Temperature annealing:} $\tau \leftarrow \max(\tau_{\mathrm{min}}, \alpha_{\tau}\tau)$
    \STATE $t \leftarrow t+1$

\ENDWHILE

\RETURN $\theta^{*}\leftarrow\theta_t,\ \psi^{*}\leftarrow\psi_t$
\end{algorithmic}
\end{algorithm}

\section{Experiments}
\subsection{Dataset}

We evaluate the proposed framework on three histopathology datasets: Dataset I, the Extended CRC dataset~\cite{extend}; Dataset II, the colon cancer dataset~\cite{china}; and Dataset III, the melanoma dataset~\cite{melanoma}. Dataset I, from the University of Warwick, contains $300$ H\&E-stained histology images extracted at $20\times$ magnification. The image resolutions are approximately $5000 \times 7300$ and $4548 \times 7520$ pixels. It contains three diagnostic categories: $120$ normal, $120$ low-grade, and $60$ high-grade images. Dataset II contains $717$ ROIs from Zhejiang University, including $355$ cancer images and $362$ normal images. The images are acquired at $40\times$ magnification, with an average spatial resolution of approximately $10000 \times 10000$ pixels. Dataset III, from NYU Langone Health, consists of $153$ H\&E-stained WSIs of metastatic melanoma tissue from $72$ patient cases. The slides are digitized at either $20\times$ or $40\times$, and all $40\times$ WSIs are downsampled to $20\times$ for consistency. The dataset includes $1072$ pathologist-annotated ROIs with an
average resolution of $4190\times4240$ pixels, comprising $544$ lymphocyte-rich and $528$ tumor-rich regions.

\subsection{Implementation}

For patch extraction, we evaluate patch sizes of $512\times512$,
$768\times768$, and $1024\times1024$ pixels with a fixed stride
of $256$ pixels. Based on the grid search, Datasets I and II are
partitioned into overlapping patches of size $768\times768$ pixels,
whereas Dataset III uses $512\times512$-pixel patches.

For Dataset I, we adopt the three-fold cross-validation scheme established in prior studies~\cite{hatnet,extend}. In each run, one fold is held out for testing, while the remaining two folds are partitioned into $80\%$ training and $20\%$ validation sets. Performance metrics are averaged across the three folds. For Dataset II, a subset of normal images is excluded because
they contain too few cells to construct valid cell graphs. This
leaves $669$ images ($355$ cancer and $314$ normal) for
evaluation. After exclusion, we apply five-fold cross-validation, holding out one
fold for testing and splitting the remaining four folds into
$80\%$ training and $20\%$ validation sets. For Dataset III, we divide the ROIs into five patient-disjoint folds. Following a similar evaluation protocol, one fold is held out for testing while the remaining four are split into $80\%$ training and $20\%$ validation sets, with final results averaged across all five folds.

The proposed framework is implemented in PyTorch. For patch-level cell graphs, we follow previous studies~\cite{hatnet,cgcnet} and set the neighborhood threshold to $r_p=64$ pixels for all datasets. Both the classifier and the adjacency generator are optimized using Adam, with learning rates selected through grid search. For Dataset I, their learning rates are $10^{-3}$ and $10^{-2}$, respectively; for Datasets II and III, they are $10^{-4}$ and $10^{-3}$. The unrolled inner-step learning rate is set to $10^{-4}$, and the finite-difference radius is set to $10^{-2}$. We use a batch size of $20$ for all experiments.  The Gumbel-Sigmoid temperature is initialized at $1.0$ to
encourage early exploration, annealed by a factor of $0.98$,
and bounded below at $0.1$ to retain moderate stochasticity. 
Training is performed until validation convergence, and we use the same model architecture for training all three datasets. The combined classifier and adjacency generator contain approximately $0.86M$ trainable parameters.

\subsection{Experimental Results}

To evaluate the effectiveness of the proposed framework, we compare $G_2^*$-Net with representative CNN- and graph-based methods, as summarized in Table~\ref{tab:consolidated_results_v3}. Overall, our method achieves the best performance on Datasets I and III and remains competitive on Dataset II.

On Dataset I, our method achieves an accuracy of $96.33 \pm 0.58\%$, outperforming all baselines. It improves over the
best CNN baseline, Xception, by $9.66\%$ and
the best GCN baseline, HAT-Net, by $1.00\%$. On Dataset II, $G_2^*$-Net achieves $97.03 \pm 1.68\%$ accuracy. Although CGC-Net obtains the highest accuracy on this dataset, the proposed method remains competitive with both CNN-based and graph-based baselines. 
On Dataset III, our method achieves the best performance with an accuracy of $97.29 \pm 0.89\%$. It outperforms the established CNN- and GCN-based approaches, demonstrating that the proposed framework is not limited to colorectal cancer analysis. The performance gain on this dataset suggests that our proposed method can generalize to a different histopathology classification setting involving melanoma tissue.

\begin{table*}[h]
\centering
\caption{Performance comparison on Dataset I, II and III. Results are reported as mean $\pm$ standard deviation}
\label{tab:consolidated_results_v3}
\setlength{\tabcolsep}{4pt}
\begin{tabular}{|p{2.5cm}|p{3.4cm}|p{3.4cm}|p{3.4cm}|}
\hline
\centering\textbf{Method} 
& \centering\textbf{Dataset I}\\ \textbf{Extended CRC} \textbf{(\%)}\\ 
& \centering\textbf{Dataset II}\\ \textbf{Colon Cancer} \textbf{(\%)}\\  
& \centering\textbf{Dataset III}\\ \textbf{Melanoma} \textbf{(\%)}\\  \tabularnewline
\hline

ResNet50 \cite{resnet50} 
& \centering $86.33 \pm 0.94$ 
& \centering $96.52 \pm 0.50$ 
& \centering $94.78 \pm 2.40$ \tabularnewline \hline

MobileNet \cite{mobilenet} 
& \centering $84.33 \pm 3.30$ 
& \centering $94.36 \pm 2.07$ 
& \centering $92.64 \pm 2.79$ \tabularnewline \hline

InceptionV3 \cite{inceptionv3} 
& \centering $84.67 \pm 1.70$ 
& \centering $92.04 \pm 2.98$ 
& \centering $96.08 \pm 1.88$ \tabularnewline \hline

Xception \cite{xception} 
& \centering $86.67 \pm 0.94$ 
& \centering $96.68 \pm 0.76$ 
& \centering $93.19 \pm 2.31$ \tabularnewline \hline


ViT \cite{vit} 
& \centering $86.67 \pm 4.04$ 
& \centering $--$ 
& \centering $--$ \tabularnewline \hline

CGC-Net \cite{cgcnet} 
& \centering $93.00 \pm 0.93$ 
& \centering $\mathbf{97.68 \pm 1.04}$ 
& \centering $95.06 \pm 1.30$ \tabularnewline \hline

HAT-Net \cite{hatnet} 
& \centering $95.33 \pm 0.58$ 
& \centering $--$ 
& \centering $96.67 \pm 1.53$ \tabularnewline \hline

C2P-GCN \cite{c2pgcn} 
& \centering $95.00 \pm 1.70$ 
& \centering $96.60 \pm 2.09$ 
& \centering $95.15 \pm 1.16$ \tabularnewline \hline

\textbf{$G_2^*$-Net (Ours)} 
& \centering $\mathbf{96.33 \pm 0.58}$ 
& \centering ${97.03 \pm 1.68}$ 
& \centering $\mathbf{97.29 \pm 0.89}$ \tabularnewline
\hline

\end{tabular}
\end{table*}

\subsection{Computational Cost of Image-Level Graph Learning}

The learned image-level graph in our framework requires pairwise edge scoring between patch nodes, leading to $\mathcal{O}(n_i^2)$ complexity for an image with $n_i$ patches. In the full-grid setting, the estimated node counts are $486$, $1369$, and $256$ for Datasets I, II, and III, respectively. A dense \texttt{float32} adjacency matrix requires $M_A(n_i)=4n_i^2/10^6~\mathrm{MB}$, corresponding to
$0.94$, $7.50$, and $0.26~\mathrm{MB}$ per image, or $18.9$, $149.9$, and $5.2~\mathrm{MB}$ for a batch of $20$. Although training retains additional tensors, including edge logits and sampled and symmetrized adjacency matrices, the memory requirement
remains feasible. Thus, image-level graph learning is computationally practical for the graph sizes considered in this study.

\section{Ablation study}

In this section, we evaluate image-level connectivity and patch-level feature families through ablation studies on Dataset I, whose
multiclass setting provides a challenging benchmark.

\subsection{Impact of Learned Image-Level Connectivity}

To evaluate learned image-level connectivity, we replace the learned adjacency with a fixed graph based on cosine similarity between patch descriptors, while keeping all other components
and training settings unchanged. As shown in
Table~\ref{tab:ablation_fixed_graph}, the fixed graph reduces accuracy from $96.33 \pm 0.58\%$ to $95.00 \pm 1.70\%$, supporting the benefit of bilevel connectivity learning over predefined similarity.

\begin{table}[h]
\centering
\caption{Image-level adjacency ablation on Dataset I}
\label{tab:ablation_fixed_graph}
\begin{tabular}{|l|c|}
\hline
\textbf{Approach} & \textbf{Accuracy (\%)} \\
\hline
$G_2^*$-Net & $96.33 \pm 0.58$ \\
\hline
Fixed image-level graph & $95.00 \pm 1.70$ \\
\hline
\end{tabular}
\end{table}

\subsection{Impact of Patch-Level Feature Groups}

To assess the importance of the patch-level feature sets, we evaluate $G_2^*$-Net using individual feature groups separately. As shown in Table~\ref{tab:feature_ablation}, all individual
groups underperform the full feature set. Nuclear nearest-neighbor
features achieve the highest individual-group accuracy, while the
combined features provide the best performance, indicating that
the feature groups capture complementary information.

\begin{table}[!h]
\centering
\caption{Patch-level feature-group ablation on Dataset I.}
\label{tab:feature_ablation}
\begin{tabular}{|l|c|}
\hline
\textbf{Patch-level Feature Group} & \textbf{Accuracy (\%)} \\
\hline
Cell graph features (CG) & $79.67 \pm 4.04$ \\
\hline
Voronoi diagram features (VD) & $85.67 \pm 6.43$ \\
\hline
Delaunay triangulation features (DT) & $85.00 \pm 8.00$ \\
\hline
Minimum spanning tree features (MST) & $81.00 \pm 7.00$ \\
\hline
Nuclear nearest-neighbor features (NN) & $92.33 \pm 2.52$ \\
\hline
Full feature set & $96.33 \pm 0.58$\\
\hline
\end{tabular}
\end{table}

\section{Conclusion}

In this work, we propose $G_2^*$-Net, an optimized two-level
graph learning framework for histopathology image classification.
The method combines fixed patch-level graphs with a learnable
image-level graph to model local cellular organization and adaptive
tissue interactions. By formulating image-level graph
learning as a second-order bilevel problem, $G_2^*$-Net separates
graph connectivity learning from classifier optimization and
incorporates second-order hypergradients via a
one-step unrolled approximation. Experiments on three
histopathology datasets demonstrate strong and competitive
performance against CNN- and GCN-based baselines.

In the future, we will extend the framework to learn both patch- and image-level graphs, enabling joint optimization of local cell connectivity and global tissue interactions, potentially yielding more informative tissue representations.

\AtNextBibliography{\footnotesize}
\printbibliography

\end{document}